\documentclass[%
 amsmath,amssymb,
 aps,prl,twocolumn,
]{revtex4-1}

\usepackage{graphicx}% Include figure files
\usepackage{dcolumn}% Align table columns on decimal point
\usepackage{bm}% bold math
\usepackage{color}% bold math
\begin{document}

\title{Chiral Phonon Mediated High-Temperature Superconductivity}% Force line breaks with \\

\author{Yi Gao}\thanks{These authors contributed equally to this work.}
\affiliation{Phonon Engineering Research Center of Jiangsu Province, Center for Quantum Transport and Thermal Energy Science, Institute of Physics Frontiers and Interdisciplinary Sciences, School of Physics and Technology, Nanjing Normal University, Nanjing 210023, China}

\author{Yang Pan}\thanks{These authors contributed equally to this work.}
 \affiliation{Phonon Engineering Research Center of Jiangsu Province, Center for Quantum Transport and Thermal Energy Science, Institute of Physics Frontiers and Interdisciplinary Sciences, School of Physics and Technology, Nanjing Normal University, Nanjing 210023, China}%Lines break automatically or can be forced with \\

\author{Jun Zhou}%
\email{zhoujunzhou@njnu.edu.cn}
\affiliation{Phonon Engineering Research Center of Jiangsu Province, Center for Quantum Transport and Thermal Energy Science, Institute of Physics Frontiers and Interdisciplinary Sciences, School of Physics and Technology, Nanjing Normal University, Nanjing 210023, China}%

\author{Lifa Zhang}
\email{phyzlf@njnu.edu.cn}
\affiliation{Phonon Engineering Research Center of Jiangsu Province, Center for Quantum Transport and Thermal Energy Science, Institute of Physics Frontiers and Interdisciplinary Sciences, School of Physics and Technology, Nanjing Normal University, Nanjing 210023, China}

\date{\today}% It is always \today, today,

\begin{abstract}
Breaking down the traditional perception on phonons which are achiral, the recent discovery of chiral phonon carrying angular momentum provides possible ways to couple electron, photon, spin, magnon and exciton, etc. We theoretically proposed an electron-chiral phonon interaction with two-phonon process, in contrast to conventional electron-phonon interaction, and a kind of effective Hubbard U through exchanging two chiral phonons is proposed. Taking two-dimensional diatomic honeycomb lattice as an example, we found this repulsive Hubbard U mediated by chiral phonons induces unconventional and high-temperature superconductivity. Moreover, the numerical calculations show an inverse isotope effect which is consistent with experimental observations in high-$T_c$ superconductors. Our finding on electron-chiral phonon and the associated Cooper pair provides a path to understand the high-$T_c$ superconductivity.
%\begin{description}
%use the optional argument of the \verb+\item+ command to give the category of %each item.
%\end{description}
\end{abstract}

\maketitle

%\tableofcontents
 The electron-phonon (E-P) interaction is not only the key to conventional
 superconductivity according to the Bardeen-Cooper-Schrieffer (BCS) theory, but also an important factor of the mechanism of
 high transition temperature ($T_{c}$) superconductivity \cite{KresinRMP2009}. Pairs of bound electrons, also called ``Cooper pairs'', can be formed through
 exchanging intermediate phonons which leads to an effective attraction
 between electrons. It was still under debate that whether the E-P interaction is adequate to explain the high-$T_{c}$ superconductivity in cuprates because of the peculiar phenomena including: the weak or inverse isotope effect, the linear electrical resistivity, and the kink in the electronic dispersion relations \cite{LanzaraNature2001,GiustinoNature2008}, etc. Other E-P interaction effects such as the unconvetional E-P interaction \cite{SCbook}, the polaronic effect \cite{KresinRMP2009}, the non-adiabatic effect \cite{SCbook1}, spatial charge inhomogeneity \cite{ReznikNature2006}, and the rapid increase of E-P coupling strength below the critical point in optimally doped strange metal \cite{HeScience2018} further complicate the underlying mechanisms.
 Beside phonons, other bosons like magnons and plasmons were proposed to be responsible for the formation of Cooper pairs in doped Mott insulator \cite{LeeRMP2006}.

 Recently, a nontrivial chiral phonon effect, which
 characterizes the phonon angular momentum (PAM) \cite{McLellanJPC1988}, has been theoretically proposed by Zhang et al. \cite{ZhangPRL2014,ZhangPRL2015} and has been experimentally observed in both  two-dimensional materials \cite{ZhuScience2018} and simple ferromagnets \cite{TauchertNature2022}. Interestingly, the chiral phonon effect
 was also observed through the measurement of thermal Hall conductivity in the pseudogap phase of cuprates \cite{GrissNP2020} which are typical high-$T_c$ superconductors. This finding could be attributed to the spin-phonon interaction \cite{StammNM2007,TauchertNature2022}. It is straightforward to speculate that the PAM would strongly affect the interplay of spin, charge and phonon in high-$T_{c}$ superconductors and consequently modify the superconducting (SC) phase transition.

 This chiral phonon effect cannot be simply incorporated into the BCS theory. Because the key quantities of conventional E-P interaction in BCS theory are phonon frequency $\omega$, phonon density-of-states (DOS), and E-P coupling constant. The PAM is characterized by ${\bf L}^{\text{ph}}_{l,s}={\bf u}_{l,s}\times {\bf p}_{l,s}$ which describes the rotation of ions around their equilibrium
 positions \cite{ZhangPRL2014}. ${\bf u}_{l,s}$ and ${\bf p}_{l,s}$ are the displacement and momentum of $s$-th atom in $l$-th unit cell,
 respectively. Usually, the overall PAM ${\bf
   L}^{\text{ph}}=\sum_{_{l,s}}{\bf L}^{\text{ph}}_{l,s}$ vanishes
 due to the time-reversal and inversion symmetries. A nonzero ${\bf
   L}^{\text{ph}}$ and the corresponding phonon magnetization (${\bf M}^{\text{ph}}$) can be obtained in nonequilibrium chiral system where both time-reversal and
 inversion symmetries are broken \cite{HamadaPRL2018,HamadaPRR2020}.

In this Letter, we consider the electron-chiral phonon (E-CP) interaction in
the framework of strong-coupling theory in the Hubbard model
\cite{TakimotoPRB2004}. This interaction is non-adiabatic and is beyond the
Born-Oppenheimer approximation because ${\bf
   L}^{\text{ph}}_{l,s}$ is a function of both ${\bf u}_{l,s}$
and ${\bf p}_{l,s}$. In other words, the electronic motion does not only depend on ion coordinations but also on their momenta. Our
numerical calculations reveal the significant contribution of
E-CP interaction to superconductivity.

{\sl Method.}-- The Hamiltonian is consisted of four parts. $H=H_0+H_{ee}+H_{ep}+H_{e-cp}$ where $H_0$ is the noninteracting part, $H_{ee}$ is the electron-electron interaction,  $H_{ep}$ is the conventional E-P interaction, and $H_{e-cp}$ describes the E-CP interaction. We choose the Planck constant $\hbar$, Boltzmann constant $k_{B}$ and lattice constant to be 1 for simplicity. We adopt a tight-binding model for a two-dimensional (2D) diatomic honeycomb lattice, then
%\begin{equation}
$H_0=\sum_{\mathbf{k}\sigma}\psi_{\mathbf{k}\sigma}^{\dag}M_{\mathbf{k}}\psi_{\mathbf{k}\sigma},$
%\end{equation}
where $\psi_{\mathbf{k}\sigma}^{\dag}=(c_{\mathbf{k}1\sigma}^{\dag},c_{\mathbf{k}2\sigma}^{\dag})$ and $\psi_{\mathbf{k}\sigma}=(c_{\mathbf{k}1\sigma},c_{\mathbf{k}2\sigma})^{T}$. $c_{\mathbf{k}s\sigma}^{\dag}$ and $c_{\mathbf{k}s\sigma}$ are creation and annihilation operators of electrons with momentum $\mathbf{k}$ and spin $\sigma$ ($\sigma=\uparrow,\downarrow$) on sublattice $s$ ($s=1,2$).
\begin{eqnarray}
\label{h0}
M_{\mathbf{k}}&=&\begin{pmatrix}
-\mu&\varepsilon_{\mathbf{k}}\\
\varepsilon_{\mathbf{k}}^{*}&-\mu
\end{pmatrix},
\end{eqnarray}
where
%\begin{eqnarray}
%\label{ek}
$
\varepsilon_\mathbf{k}=t[e^{i(\frac{3}{2}k_x+\frac{\sqrt{3}}{2}k_y)}+e^{i\sqrt{3}k_y}+e^{i(\frac{3}{2}k_x+\frac{3\sqrt{3}}{2}k_y)}],
$
%\end{eqnarray}
$t$ is the nearest-neighbor hopping integral, and the electron wave vector is ${\mathbf k}=({k_x,k_y})$.

The electron-electron interaction is considered as the Hubbard type, which is
%\begin{eqnarray}
%\label{Hee}
$H_{ee}=\frac{U}{N}\sum_{\mathbf{k},\mathbf{k}^{'},\mathbf{q},s}c_{\mathbf{k}+\mathbf{q}s\uparrow}^{\dagger}c_{\mathbf{k}^{'}-\mathbf{q}s\downarrow}^{\dagger}c_{\mathbf{k}^{'}s\downarrow}c_{\mathbf{k}s\uparrow},$
%\end{eqnarray}
where $N$ is the number of unit cells and $U$ is the Coulomb interaction. The conventional E-P interaction is written as 
$H_{ep}=\frac{\tilde{g}_{ep}}{\sqrt{N}}\sum_{\mathbf{q},\mathbf{k},s,\sigma,\nu}c^{\dag}_{\mathbf{k}+\mathbf{q}s\sigma}c_{\mathbf{k}s\sigma}A^{\nu}_{\mathbf{q}}$, where $\tilde{g}_{ep}=g_{ep}m^{-1/4}$ is the E-P coupling strength and $m$ is the mass of atom. $A_{\mathbf{q}}^{\nu}=a_{\mathbf{q}}^{\nu}+a_{-\mathbf{q}}^{\nu\dag}$ where $a_{\mathbf{q}}^{\nu\dagger}$ ($a_{\mathbf{q}}^{\nu}$) creates (annihilates) a phonon with wave vector $\mathbf{q}$ and mode $\nu$.

We assume that electron spin ($\mathbf{S}_{l,s}$) feels a magnetic field induced by the local atomic rotation where the magnetic field is proportional to the PAM. Then the E-CP interaction is analogous to the form of spin-orbit coupling $\eta \mathbf{L}_{l,s} \cdot \mathbf{S}_{l,s}$ \cite{Shankarbook} where $\eta$ is the coupling strength parameter. In order to simplify our calculations, we assume that $\eta$ is a constant and $L_{l,s}^{x}S_{l,s}^{x}$ and $L_{l,s}^{y}S_{l,s}^{y}$ are negligible. The Hamiltonian of E-CP interaction is written as
\begin{eqnarray}
\label{Hep}
H_{e-cp}&=&2\eta\sum_{l,s}L_{l,s}^{z}S_{l,s}^{z}\nonumber\\
&=&-\frac{\sqrt{2}\eta}{N}\sum_{\nu,\nu^{'},\mathbf{q},\mathbf{q}^{'},s}g^{\nu,\nu^{'}}_{\mathbf{q},\mathbf{q}^{'},s}A_{\mathbf{q}}^{\nu}B_{-(\mathbf{q}-\mathbf{q}^{'})}^{\nu^{'}}S_{\mathbf{q}^{'}}^{zss},\nonumber\\
\end{eqnarray}
where $B_{\mathbf{q}}^{\nu}=a_{\mathbf{q}}^{\nu}-a_{-\mathbf{q}}^{\nu\dag}$ and $S_{\mathbf{q}}^{zss}=\frac{1}{\sqrt{2}}\sum_{\mathbf{k},\sigma}\sigma c_{\mathbf{k}+\mathbf{q}s\sigma}^{\dag}c_{\mathbf{k}s\sigma}$. Detailed derivation of $H_{e-cp}$ is given in Supplemental Material. The matrix element is defined as
\begin{eqnarray}
g^{\nu,\nu^{'}}_{\mathbf{q},\mathbf{q}^{'},s}&=&\sqrt{\frac{\omega_{\mathbf{q}-\mathbf{q}^{'}}^{\nu^{'}}}{\omega_{\mathbf{q}}^{\nu}}}\xi_{\mathbf{q}-\mathbf{q}^{'},\nu^{'}}^{\dag}(s)
\begin{pmatrix}
0&-i\\
i&0
\end{pmatrix}
\xi_{\mathbf{q},\nu}(s),
\end{eqnarray}
where $\xi_{\mathbf{q},\nu}(s)$ is the polarization vector and $\omega_{\mathbf{q}}^{\nu}$ is the phonon frequency. It satisfies
$g^{\nu,\nu^{'}}_{-\mathbf{q},-\mathbf{q}^{'},s}=-g^{\nu,\nu^{'}*}_{\mathbf{q},\mathbf{q}^{'},s}$.

\begin{figure}
\includegraphics[width=1\linewidth]{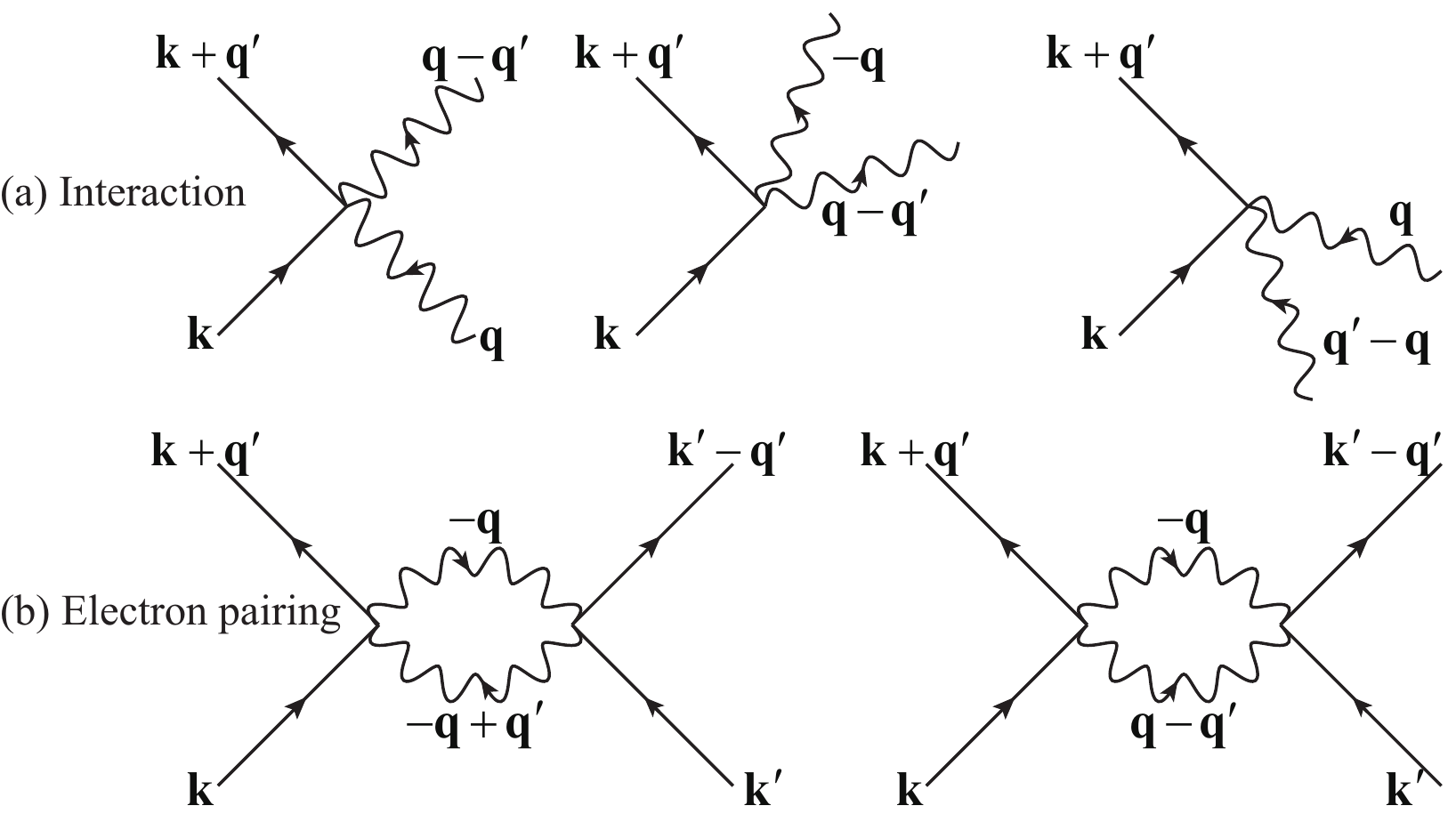}
 \caption{\label{feynman diagram} The Feynman diagrams of E-CP interaction. (a) Three cases of E-CP interaction for one electron and two phonons. (b) Two cases in which pairs are formed by exchanging two phonons.  }
\end{figure}

The Feynman diagrams of E-CP interaction are shown in Fig. \ref{feynman diagram}. This interaction is non-adiabatic, in which electrons are affected by both the displacement and momentum of the atoms. The phonon operators in the E-CP interaction are $A_{\bf{q}}^\nu B_{ - \left( {{\bf{q}} - {\bf{q'}}} \right)}^{\nu '}$ which can be expanded as
\begin{eqnarray}
	 \left( { - a_{\bf{q}}^\nu a_{{\bf{q}} - {\bf{q'}}}^{\nu '\dag } + a_{ - {\bf{q}}}^{\nu \dag }a_{{\bf{q'}} - {\bf{q}}}^{\nu '}} \right) - a_{ - {\bf{q}}}^{\nu \dag }a_{{\bf{q}} - {\bf{q'}}}^{\nu '\dag } + a_{\bf{q}}^\nu a_{{\bf{q'}} - {\bf{q}}}^{\nu '}.
\end{eqnarray}
So there are three kinds of two-phonon processes: emitting one phonon and absorbing one phonon; emitting two phonons; absorbing two phonons, which are shown in Fig. \ref{feynman diagram} (a). Consequently, the second-order of the aforementioned two-phonon processes give rise to two kinds of effective electron-electron interactions as shown in Fig. \ref{feynman diagram} (b).
Unlike the BCS mechanism in which Cooper pairs are formed through exchanging one phonon, the E-CP interaction gives another mechanism of the formation of Cooper pairs via en effective Hubbard U induced through exchanging two chiral phonons. The Hamiltonian of the effective electron-electron interaction is given in Supplemental Material. We must emphasize that the E-CP interaction is fundamentally different from the conventional anharmonic E-P interaction with two-phonon absorption and emission. The chiral phonons are able to affect the spin fluctuation and the magnetic order which can not be affected by anharmonic E-P interaction.

We adopt the widely used Green's function method to investigate the effect of E-CP interaction on high-$T_c$ superconductivity. First we calculate the electron's normal Green's function matrix $G(k)$ according to Eq. (\ref{h0}). Then the irreducible susceptibility matrix $\chi_0(q)$ can be calculated based on $G(k)$. Then within the random phase approximation (RPA), we can calculate the longitudinal and transverse spin susceptibility matrices, $\chi_{s}^{zz}(q)$ and $\chi_{s}^{+-}(q)$, respectively, as well as the charge susceptibility matrix, $\chi_{c}(q)$.

When the temperature is close to $T_c$, the linearized Eliashberg equation can be expressed as \cite{TakimotoPRB2004}
\begin{eqnarray}
\label{Eliashberg}
\lambda \phi^{a_6a_5}(k)&=&-\frac{T}{N}\sum_q\sum_{a_1,a_2,a_3,a_4} G^{a_1a_2}(k-q)G^{a_4a_3}(q-k)\nonumber\\
&&\phi^{a_2a_3}(k-q)V^{a_4a_5,a_1a_6}(q),
\end{eqnarray}
where $\phi(k)$ is the electron's anomalous self-energy. $T_c$ is reached when the largest eigenvalue $\lambda$ in Eq. (\ref{Eliashberg}) reaches $1$. The singlet pairing interaction in Eq. (\ref{Eliashberg}) is \cite{TakimotoPRB2004}
\begin{eqnarray}
\label{V}
V(q)&=&\frac{1}{2}\tilde{S}(q)\chi_{s}^{zz}(q)\tilde{S}(q)+U_s\chi_{s}^{+-}(q)U_s\nonumber\\
&&-\frac{1}{2}\tilde{C}(q)\chi_{c}(q)\tilde{C}(q)+\frac{1}{2}[\tilde{S}(q)+\tilde{C}(q)],
\end{eqnarray}
where
\begin{widetext}
\begin{eqnarray}
\label{S}
\tilde{S}^{a_1a_1,a_2a_2}(q)&=&U\delta_{a_1a_2}-\frac{\eta^2}{N}\sum_{\mathbf{q}_1,\nu_1,\nu_1^{'}}\frac{2(N_1-N_2)(\omega_1-\omega_2)}{(\omega_1-\omega_2)^2+\omega_n^2}(g^{\nu_1,\nu_1^{'}}_{\mathbf{q}_1,-\mathbf{q},a_1} g^{\nu_1^{'},\nu_1}_{\mathbf{q}_1+\mathbf{q},\mathbf{q},a_2}+g^{\nu_1^{'},\nu_1*}_{\mathbf{q}_1+\mathbf{q},\mathbf{q},a_1}g^{\nu_1,\nu_1^{'}*}_{\mathbf{q}_1,-\mathbf{q},a_2}+2g^{\nu_1,\nu_1^{'}}_{\mathbf{q}_1,-\mathbf{q},a_1}g^{\nu_1,\nu_1^{'}*}_{\mathbf{q}_1,-\mathbf{q},a_2})\nonumber\\
&-&\frac{\eta^2}{N}\sum_{\mathbf{q}_1,\nu_1,\nu_1^{'}}\frac{2(N_1+N_2+1)(\omega_1+\omega_2)}{(\omega_1+\omega_2)^2+\omega_n^2}(g^{\nu_1,\nu_1^{'}}_{\mathbf{q}_1,-\mathbf{q},a_1} g^{\nu_1^{'},\nu_1}_{\mathbf{q}_1+\mathbf{q},\mathbf{q},a_2}+g^{\nu_1^{'},\nu_1*}_{\mathbf{q}_1+\mathbf{q},\mathbf{q},a_1}g^{\nu_1,\nu_1^{'}*}_{\mathbf{q}_1,-\mathbf{q},a_2}-2g^{\nu_1,\nu_1^{'}}_{\mathbf{q}_1,-\mathbf{q},a_1}g^{\nu_1,\nu_1^{'}*}_{\mathbf{q}_1,-\mathbf{q},a_2}).\nonumber\\
\end{eqnarray}
\end{widetext}
Here $\omega_1=\omega^{\nu_1}_{\mathbf{q}_1}$, $\omega_2=\omega^{\nu_1^{'}}_{\mathbf{q}_1+\mathbf{q}}$ are the phonon dispersion relations and $N_1=\frac{1}{e^{\omega_1/T}-1}$, $N_2=\frac{1}{e^{\omega_2/T}-1}$ are the Bose distribution functions. $\omega_{n}=2n\pi T$ is the Matsubara frequency where $n$ is integer. The derivations and expressions of $G(k)$, $\chi_0(q)$, $\chi_{s}^{zz}(q)$, $\chi_{s}^{+-}(q)$, $\chi_{c}(q)$, $U_s$ and $\tilde{C}(q)$ can all be found in the Supplemental Material.

In the following, we set $U=2.3t$ and adjust the chemical potential $\mu$ to ensure the electron filling $n_f=0.95$ where $n_f=1+T{N}^{-1}\sum_{a_1}\sum_{k}G^{a_1a_1}(k)$.
The band width of Eq. (\ref{h0}) is about $6t$, therefore this choice of $U$ corresponds to an intermediate strength of electron-electron interaction. Furthermore since the band structure has particle-hole symmetry, we consider only the hole-doped case and the choice of $n_f$ corresponds to $5\%$ hole doping. The number of unit cells is set to be $N=32\times32$ and we use $16384$ Matsubara frequencies.
 %($-16384\pi T\leq\omega_n\leq16384\pi T$ and $-16383\pi T\leq p_n\leq16383\pi T$).
 The summation over momentum and frequency are both done by fast Fourier transformation.
 The E-CP coupling matrix element $g^{\nu,\nu^{'}}_{\mathbf{q},\mathbf{q}^{'},s}$ is calculated by given values of $m_1$ and $m_2$ which are the masses of two atoms, respectively. When $m_1=m_2=1$, we rescale the phonon's dispersion so the maximal $\omega_{\mathbf{q}}^{\nu}\approx0.097t$. The exact calculation of E-CP coupling strength $\eta$ is not an easy task. We speculate that the E-CP interaction is similar to the spin-orbit coupling by simply considering an ion rotating around an electron \cite{Shankarbook}. Then the value of $\eta$ must decrease with increasing temperature because of the enlarged distance between electron and ion. Therefore, we assume a simple exponential decay of $\eta$ as follows
\begin{eqnarray}
\label{eta}
\eta=\eta_0 e^{-T/T^{*}}.
\end{eqnarray}
Here we set $\eta_0=4.6574\times 10^{-4}t$ and $T^{*}=0.02t$.

\begin{figure}
\includegraphics[width=1\linewidth]{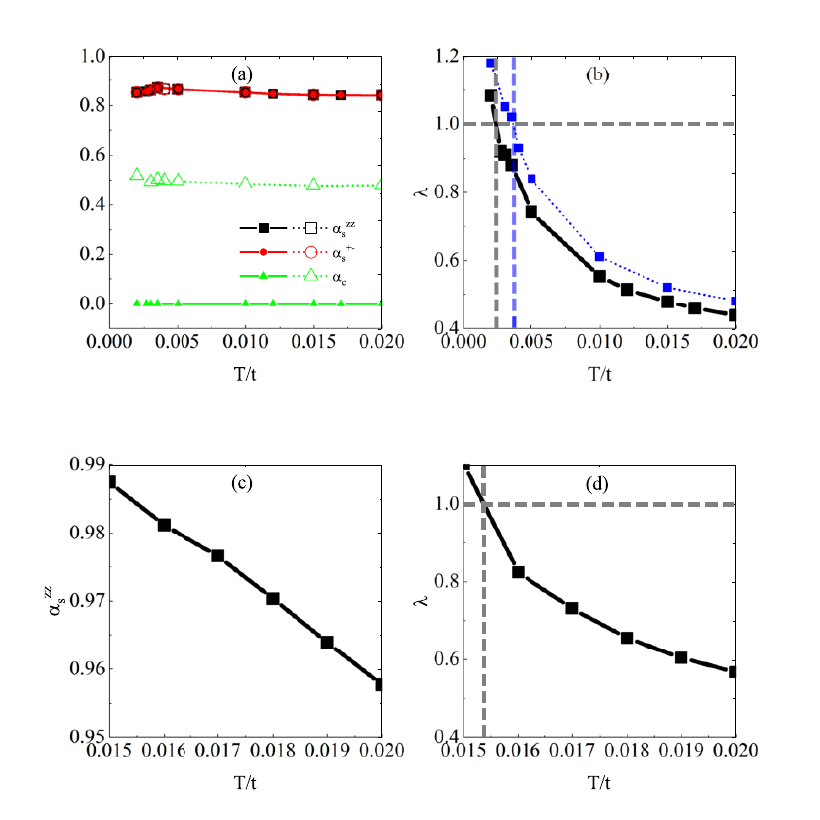}
 \caption{\label{susceptibility&lambda} (a) The evolution of $\alpha_s^{zz}$, $\alpha_s^{+-}$ and $\alpha_c$ with temperature when $\eta=\tilde{g}_{ep}=0$ (solid curves) and $\eta=0, \tilde{g}_{ep}=0.15t$ (dotted curves). (b) $\lambda$ as a function of $T$ at $\eta=\tilde{g}_{ep}=0$ (solid curve) and $\eta=0, \tilde{g}_{ep}=0.15t$ (dotted curves). (c) The evolution of $\alpha_s^{zz}$ and (d) the evaluation of $\lambda$ with temperature when $\eta$ is set according to Eq. (\ref{eta}) and $\tilde{g}_{ep}=0$. The vertical dashed lines in (b) and (d) denote the value of $T$ when $\lambda=1$. $U=2.3t$ is used in all the calculations.}
\end{figure}

\begin{figure}
\includegraphics[width=1\linewidth]{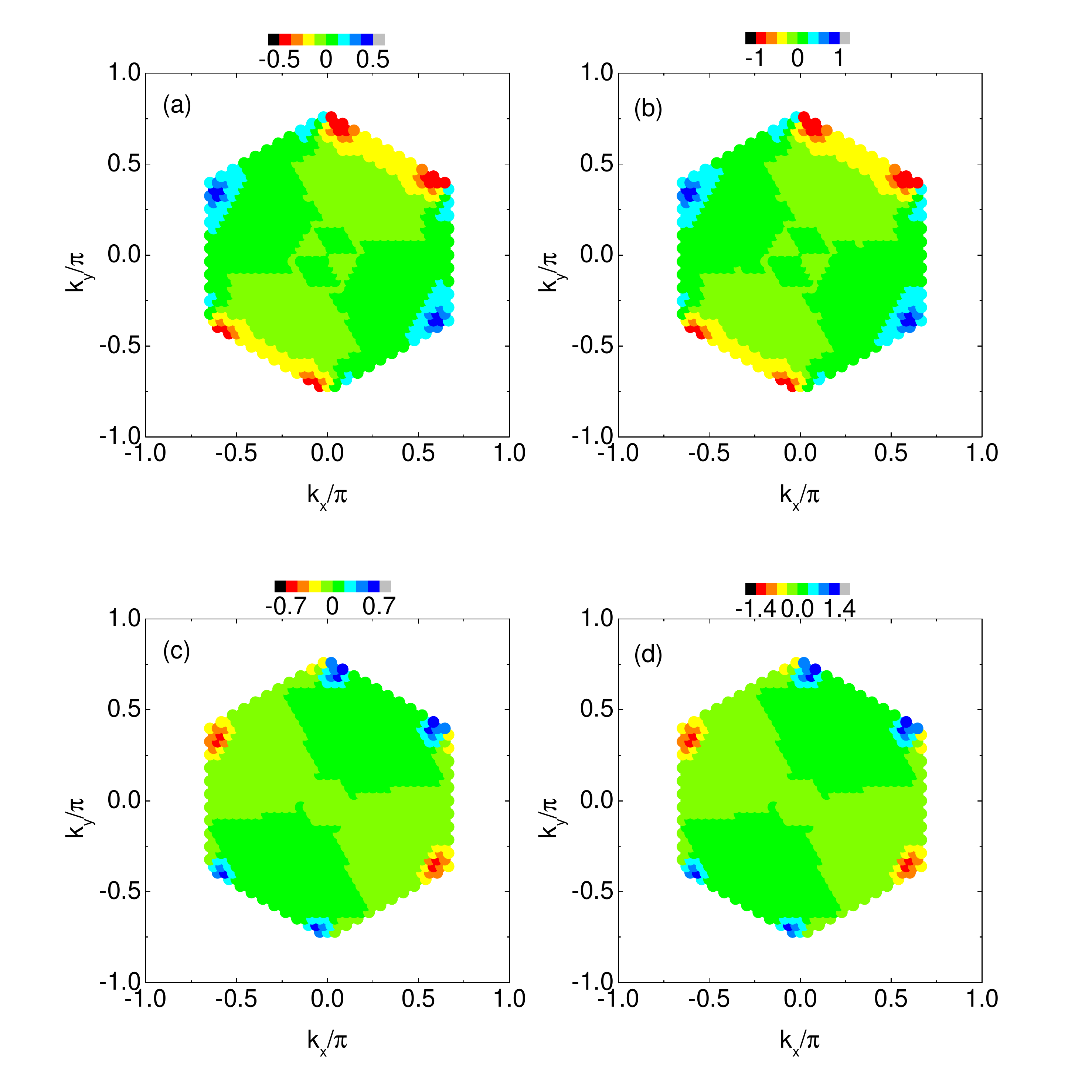}
 \caption{\label{pairing symmetry eta=0} (Color online) The calculated (a) real part $Re\Delta^{11}(\mathbf{k},i\pi T)$ and (b) imaginary part $Im\Delta^{11}(\mathbf{k},i\pi T)$ of paring function $\Delta^{11}(\mathbf{k},i\pi T)$ in the Brillouin zone when $\eta=\tilde{g}_{ep}=0$ and $T=0.002t$. (c) and (d) are similar to (a) and (b), respectively, but at $\eta\neq0$, $\tilde{g}_{ep}=0$} and $T=0.015t$.
\end{figure}

\begin{figure}
\includegraphics[width=0.7\linewidth]{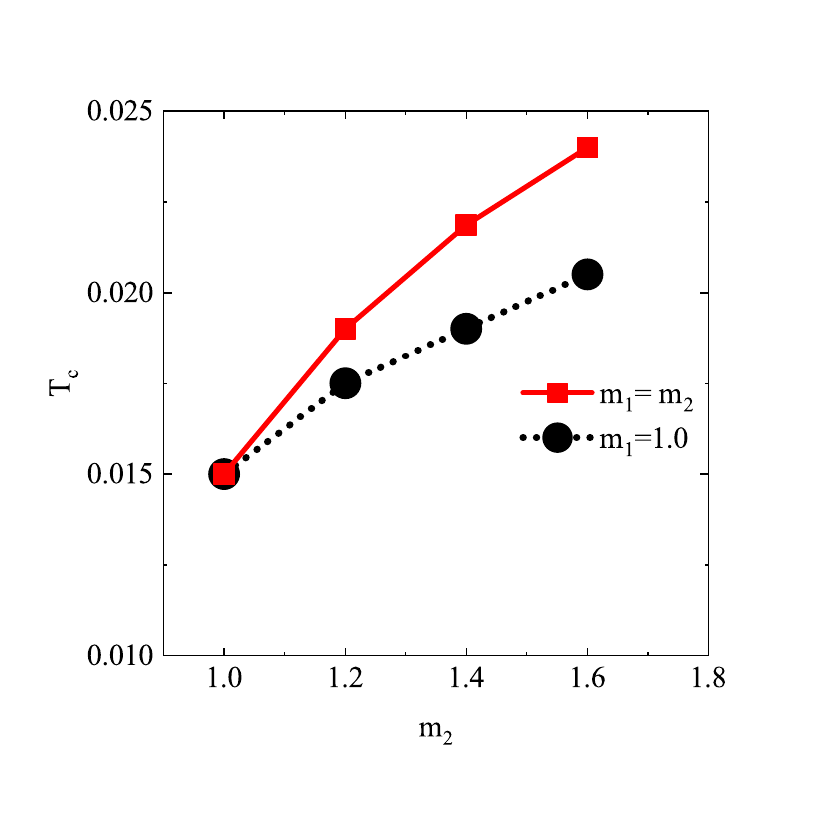}
 \caption{\label{isotope} The isotope effect of $T_c$. The calculated values of $T_c$ versus $m_2$ when $m_1=m_2=m$ and $m_1=1$. Here $\tilde{g}_{ep}=0$.}
\end{figure}

{\sl Results and discussions.} -- Starting from the electron filling $n_f=0.95$, we calculate the largest eigenvalue of $\chi_{0}(\mathbf{q},i\omega_n=0)\tilde{S}(\mathbf{q},i\omega_n=0)$ over all $\mathbf{q}$ and denote it as $\alpha_s^{zz}$. Its value has to be less than $1$ to prevent the formation of static magnetic order. We further denote $\alpha_s^{+-}$ and $\alpha_c$ as the largest eigenvalues of $\chi_{0}(\mathbf{q},i\omega_n=0)U_s$ and $-\chi_{0}(\mathbf{q},i\omega_n=0)\tilde{C}(\mathbf{q},i\omega_n=0)$, respectively. Similarly, both of them have to be less than $1$ for the system to stay away from static magnetic and charge ordering.

In the absence of E-CP and E-P couplings, i.e., $\eta=\tilde{g}_{ep}=0$, we show the evolution of $\alpha_s^{zz}$, $\alpha_s^{+-}$ and $\alpha_c$ with temperature in solid curves in Fig. \ref{susceptibility&lambda}(a). It is obvious that $\alpha_s^{zz}=\alpha_s^{+-}$. Through the temperature range we investigated, $\alpha_s^{zz}$, $\alpha_s^{+-}$ and $\alpha_c$ are always less than $1$. On the other hand, in Fig. \ref{susceptibility&lambda}(b) we show the largest eigenvalue of Eq. (\ref{Eliashberg}), $\lambda$, as a function of $T$. We can see that $\lambda$ increases with decreasing $T$ and reaches $1$ at $T\approx0.0025t$, suggesting that $T_c\approx0.0025t$ when there is neither E-CP nor E-P interaction. If the E-P coupling is present, for example, $\eta=0$ and $\tilde{g}_{ep}=0.15t$, dotted curves in Figs. \ref{susceptibility&lambda}(a) and \ref{susceptibility&lambda}(b) show that $\alpha_c$ and $T_c$ increase to $0.5$ and $0.0035t$, respectively. The spin susceptibility related $\alpha_s^{zz}$ and $\alpha_s^{+-}$ are unaffected by the E-P interaction.

In the presence of the E-CP interaction (but no E-P interaction), the evolution of $\alpha_s^{zz}$ and $\lambda$ with temperature are shown in Figs. \ref{susceptibility&lambda}(c) and \ref{susceptibility&lambda}(d), respectively.
As we can see, both $\alpha_s^{zz}$ and $\lambda$ increase with decreasing temperature, suggesting the trend to have magnetic phase transition and superconductivity phase transition. However, since when $T$ decreases, $\lambda$ reaches $1$ before $\alpha_s^{zz}$ does, thus the superconductivity phase transition actually occurs and the corresponding $T_c\approx0.0155t$ as indicated by the vertical dashed line in Fig. \ref{susceptibility&lambda}(d). Therefore, the E-CP coupling can greatly enhance $T_c$ compared to its value when $\eta=0$ as shown in Fig.\ref{susceptibility&lambda}(b).
Moreover, $\alpha_s^{+-}$ and $\alpha_c$ are not shown because only the $z$-component is considered in Eq. (\ref{Hep}), then both of them stay unchanged as shown in Fig. \ref{susceptibility&lambda}(a). Our numerical calculations further confirm that the E-P interaction, by setting $\tilde{g}_{ep}=0.15t$ in this case, hardly change the value of $T_c$. This proves that the great enhancement of $T_c$ is indeed resulted from the E-CP interaction.

This finding reveals that the E-CP interaction leads to an unambiguous increase of $\alpha_s^{zz}$. The reason for this increase is, the phonon contributions, i.e., the last two terms in Eq. (\ref{S}), are positive when $a_1=a_2$ and negligibly small when $a_1\neq a_2$. Therefore the phonon can effectively increase the value of $U$ for the longitudinal spin susceptibility, leading to an increased $\alpha_s^{zz}$, which means an increased longitudinal spin fluctuation. It then enhances the term $\frac{1}{2}\tilde{S}(q)\chi_{s}^{zz}(q)\tilde{S}(q)$ in Eq. (\ref{V}), leading to an enlarged pairing interaction, which then results in an enhanced $T_c$.

We now turn to investigate the superconductivity pairing symmetry. We project $\phi(k)$ in Eq. (\ref{Eliashberg}) onto each band and since in our hole-doped case, only the lower band crosses the Fermi level, therefore we consider only the pairing function on this band and denote it as $\Delta^{11}(k)$. When $\eta=\tilde{g}_{ep}=0$, the superconductivity pairing is resulted solely from the electron-electron interaction. In this case, at $T$ ($0.002t$) slightly below $T_c$ ($0.0025t$), we show the real and imaginary parts of $\Delta^{11}(\mathbf{k},i\pi T)$ in Figs. \ref{pairing symmetry eta=0}(a) and \ref{pairing symmetry eta=0}(b), respectively. One can see that $Re\Delta^{11}(\mathbf{k},i\pi T)$ and $Im\Delta^{11}(\mathbf{k},i\pi T)$ differ only in magnitude, but have the same structure in momentum space. Therefore the phase factor between the real and imaginary parts is global and $\Delta^{11}(\mathbf{k},i\pi T)$ can be taken as real, i.e., no time-reversal symmetry breaking. Furthermore, there exist sign changes in $\Delta^{11}(\mathbf{k},i\pi T)$, indicating that the pairing symmetry is unconventional (not isotropic $s$-wave).
When $\eta\neq 0$ and $\tilde{g}_{ep}=0$, although $T_c$ is greatly enhanced, the pairing function at $T$ ($0.015t$) slightly below $T_c$ ($0.0155t$) shown in Figs. \ref{pairing symmetry eta=0}(c) and \ref{pairing symmetry eta=0}(d) remains qualitatively the same as that for $\eta=0$. The pairing still does not break the time-reversal symmetry and is unconventional as well. By examining the phonon contribution to $\tilde{S}(q)$ we found that, the last two terms in Eq. (\ref{S}) are slightly $\mathbf{q}$-dependent, therefore the E-CP coupling will not vary the pairing symmetry significantly as compared to the $\eta=0$ case. We have also verified that, in the above two cases, including the E-P interaction at $\tilde{g}_{ep}=0.15t$ does not change the pairing function qualitatively.

Thus, by introducing the E-CP interaction, high-temperature and unconventional superconductivity can be induced in systems with relatively weak electron-electron interaction. For the present system, taking $t=1$ eV as the energy unit, the calculated superconductivity transition temperature $T_c$ can be boosted from $2.5$ meV (29K) to $15.5$ meV (180K). Furthermore, if the phonon couples to the $S^{x}$ and $S^{y}$ components of the electrons, $T_c$ may be further enhanced since the factor is $1$ in the $U_s\chi_{s}^{+-}(q)U_s$ term in Eq. (\ref{V}), instead of $\frac{1}{2}$ in the $\tilde{S}(q)\chi_{s}^{zz}(q)\tilde{S}(q)$ term. This will be studied in our future work.

The anomalous isotope effect has been found in many kinds of superconductors \cite{stritzker1972superconductivity,bornemann1991negative,schaeffer2015high,stucky2016isotope}.
The E-CP interaction provides an alternative explanation of such isotope effect.
First we consider a case where the masses of the two sublattices are equal ($m_1=m_2=m$). Fig. \ref{isotope} shows that $T_c$ increases with increasing $m$.
Such inverse isotope effect is opposite to the conventional E-P coupling in BCS theory where $T_c$ decreases with increasing $m$. Because the most significant term, $\omega_n=0$ component in Eq. (\ref{S}), can be written as
\begin{eqnarray}
& &\tilde{S}^{a_1a_1,a_2a_2}(\mathbf{q},i\omega_n=0,m,T)\nonumber\\
&=&\sqrt{m}\tilde{S}^{a_1a_1,a_2a_2}(\mathbf{q},i\omega_n=0,1,\sqrt{m}T),
\end{eqnarray}
when the phonon dispersion relation is proportional to $1/\sqrt{m}$ and $U=0$.
This means that the matrix elements of $\tilde{S}(q)$ are enhanced as $m$ increases, leading to an inverse isotope effect. Detailed discussion of isotope effect is given in Supplemental Material.
In addition, we also investigate the isotope substitution effect in Fig. \ref{isotope} where $m_1$ is fixed to be $1$ and $m_2$ changes from $1$ to $1.6$. We can see that $T_c$ also increases with increasing $m_2$. However, the increasing rate is smaller than that of Fig. \ref{isotope}.
Our calculation results provide a possible explanation of weak and inverse isotope effect in cuprate superconductors with oxygen isotope substitution (O$^{16}\rightarrow$O$^{18}$) \cite{pringle2000PRB,bornemann1991negative}.

In summary, we introduce the electron-chiral phonon interaction into the current theory of high $T_c$ superconductivity. This interaction leads to an interplay of electron, spin, and charge in superconductors. The numerical calculation results show a remarkable enhancement of $T_c$ induced by this interaction. Moreover, both unconventioal paring and peculiar weak and inverse isotope effect are found which are able to explain the experimental observations.

J.Z. would like to thank Prof. Xiangfan Xu for discussion.
This work is supported by National Natural Science Foundation of China (No. 11890703). J.Z. is also supported by the ``Shuangchuang" Doctor program of Jiangsu Province (JSSCBS20210341).

\providecommand{\noopsort}[1]{}\providecommand{\singleletter}[1]{#1}%

\end{document}

\renewcommand{\baselinestretch}{2}
	\baselineskip 24pt
	\title{Supplemental Material for ``Chiral Phonon Mediated High-Temperature Superconductivity''}%

\author{Yi Gao}\thanks{These authors contributed equally to this work.}
\affiliation{Phonon Engineering Research Center of Jiangsu Province, Center for Quantum Transport and Thermal Energy Science, Institute of Physics Frontiers and Interdisciplinary Sciences, School of Physics and Technology, Nanjing Normal University, Nanjing 210023, China}

\author{Yang Pan}\thanks{These authors contributed equally to this work.}
\affiliation{Phonon Engineering Research Center of Jiangsu Province, Center for Quantum Transport and Thermal Energy Science, Institute of Physics Frontiers and Interdisciplinary Sciences, School of Physics and Technology, Nanjing Normal University, Nanjing 210023, China}%Lines break automatically or can be forced with \\

\author{Jun Zhou}%
\email{zhoujunzhou@njnu.edu.cn}
\affiliation{Phonon Engineering Research Center of Jiangsu Province, Center for Quantum Transport and Thermal Energy Science, Institute of Physics Frontiers and Interdisciplinary Sciences, School of Physics and Technology, Nanjing Normal University, Nanjing 210023, China}%

\author{Lifa Zhang}
\email{phyzlf@njnu.edu.cn}
\affiliation{Phonon Engineering Research Center of Jiangsu Province, Center for Quantum Transport and Thermal Energy Science, Institute of Physics Frontiers and Interdisciplinary Sciences, School of Physics and Technology, Nanjing Normal University, Nanjing 210023, China}

\maketitle
\tableofcontents

\section{The E-CP interaction Hamiltonian}
The ${\bf u}_{l,s}$ , ${\bf p}_{l,s}$ and $\mathbf{S}^z_{l,s}$ can be written in the second quantization representation
  \begin{eqnarray}
  	\label{u}
{{\bf{u}}_{l,s}} &=& \sum\limits_{{\bf{q}},\nu } {\sqrt {\frac{1}{{2{m_s}N\omega _{\bf{q}}^\nu }}} } {{\bf{\xi }}_{{\bf{q}},\nu }}\left( s \right)\left( {a_{\bf{q}}^\nu  + a_{ - {\bf{q}}}^{\nu \dag }} \right){e^{i{\kern 1pt} {\bf{q}} \cdot {{\bf{R}}_l}}},\\
  	\label{p}
 {{\bf{p}}_{l,s}} &=& \sum\limits_{{\bf{q}},\nu } {\sqrt {\frac{{{m_s}\omega _{\bf{q}}^\nu }}{{2N}}} } i{\bf{\xi }}_{{\bf{q}},\nu }^ * \left( s \right)\left( {a_{\bf{q}}^{\nu \dag } - a_{ - {\bf{q}}}^\nu } \right){e^{ - i{\kern 1pt} {\bf{q}} \cdot {{\bf{R}}_l}}},\\
  	\label{sz}
  	{\bf{S}}_{l,s}^z &=& \sum\limits_\sigma  {\sigma c_{l,s,\sigma }^\dag c_{l,s,\sigma }^{}}\nonumber\\
  	 &=& \frac{1}{N}\sum\limits_{{\bf{k}},{\bf{k'}},\sigma } {\sigma c_{{\bf{k'}},s,\sigma }^\dag c_{{\bf{k}},s,\sigma }^{} * {e^{i{\kern 1pt} \left( {{\bf{k - k'}}} \right){\kern 1pt}  \cdot {{\bf{R}}_l}}}} .
  \end{eqnarray}
The $H_{e-cp}$ is
  \begin{eqnarray}
  \label{hecp-1}
	{H_{e - cp}}  &=& 2\eta \sum\limits_{l,s} {L_{l,s}^z} S_{l,s}^z\nonumber\\
	&=& 2\eta \sum\limits_{l,s,\sigma } {\left( {u_{l,s}^xp_{l,s}^y - u_{l,s}^yp_{l,s}^x} \right)\sigma c_{l,s,\sigma }^\dag {c_{l,s,\sigma }}}\nonumber \\
	  &=& 2\eta \sum\limits_{l,s} {{\bf{u}}_{l,s}^T\left( {\begin{array}{*{20}{c}}
				0&1\\
				{ - 1}&0
		\end{array}} \right){{\bf{p}}_{l,s}}\sigma c_{l,s,\sigma }^\dag {c_{l,s,\sigma }}} .
 \end{eqnarray}
Substitute Eq. (\ref{u}--\ref{sz}) into Eq. (\ref{hecp-1}) and take the transpose , we obtain
  \begin{align}
  	\label{hecp-2}
		{H_{e - cp}} =& \frac{{ \eta }}{N}\sum\limits_s {\sum\limits_{\scriptstyle{\bf{q}},{\bf{q'}}\hfill\atop
				\scriptstyle\nu ,\nu '\hfill} {\sum\limits_{{\bf{k}},\sigma } {\sqrt {\frac{{\omega _{{\bf{q}} - {\bf{q'}}}^{\nu '}}}{{\omega _{\bf{q}}^\nu }}} {\bf{\xi }}_{{\bf{q}} - {\bf{q'}},\nu '}^\dag \left( s \right)\left( {\begin{array}{*{20}{c}}
							0&{ - i}\\
							i&0
					\end{array}} \right){\bf{\xi }}_{{\bf{q}},\nu }^{}\left( s \right)\left( {a_{\bf{q}}^\nu  + a_{ - {\bf{q}}}^{\nu \dag }} \right)\left( {a_{{\bf{q}} - {\bf{q'}}}^{\nu '\dag } - a_{{\bf{q'}} - {\bf{q}}}^{\nu '}} \right)\sigma c_{{\bf{k}} + {\bf{q'}},s,\sigma }^\dag {c_{{\bf{k}},s,\sigma }}} } } \nonumber\\
		=& -\frac{{\sqrt 2 \eta }}{N}\sum\limits_{{\bf{q}},{\bf{q'}},\nu ,\nu ',s} {g_{{\bf{q}},{\bf{q'}},s}^{\nu ,\nu '}A_{\bf{q}}^\nu B_{ - \left( {{\bf{q}} - {\bf{q'}}} \right)}^{\nu '}S_{{\bf{q'}}}^{zss}} ,
\end{align}
which is the same as Eq. (2).

\section{The effective Hamiltonian}
By expending Eq. (\ref{hecp-2})
\begin{eqnarray}
	{H_{e - cp}} = \frac{{ - \sqrt 2 \eta }}{N}\sum\limits_{\nu \nu 'qq's} {g_{qq's}^{\nu \nu '}\left[ {\left( { - a_q^\nu a_{q - q'}^{\nu '\dag } + a_{ - q}^{\nu \dag }a_{ - \left( {q - q'} \right)}^{\nu '}} \right) + \left( { - a_{ - q}^{\nu \dag }a_{q - q'}^{\nu '\dag } + a_q^\nu a_{q' - q}^{\nu '}} \right)} \right]S_{q'}^{zss}},
\end{eqnarray}
we can devide the E-CP interaction Hamiltonian into two parts,
\begin{eqnarray}
	{H_{e - cp}} = {H_{e - cp - 1}}{\rm{ + }}{H_{e - cp - 2}},
\end{eqnarray}
where
\begin{eqnarray}
	{H_{e - cp - 1}}& =& \frac{{ - \sqrt 2 \eta }}{N}\sum\limits_{\nu \nu 'qq's} {g_{qq's}^{\nu \nu '}\left( { - a_q^\nu a_{q - q'}^{\nu '\dag } + a_{ - q}^{\nu \dag }a_{ - \left( {q - q'} \right)}^{\nu '}} \right)S_{q'}^{zss}} ,\nonumber\\
	{H_{e - cp - {\rm{2}}}}& = &\frac{{ - \sqrt 2 \eta }}{N}\sum\limits_{\nu \nu 'qq's} {g_{qq's}^{\nu \nu '}\left( { - a_{ - q}^{\nu \dag }a_{q - q'}^{\nu '\dag } + a_q^\nu a_{q' - q}^{\nu '}} \right)S_{q'}^{zss}} .
\end{eqnarray}

Then according to the two virtual phonon processes shown in Feynman diagrams in Fig. 1(b), we can write the interaction matrix elements of electrons by exchanging two chiral phonons,
\begin{eqnarray}
	&&\sum\limits_m {\left\langle f \right|{H_{e - cp}}\left| m \right\rangle \left\langle m \right|{H_{e - cp}}\left| i \right\rangle }/\left(E_i-E_m\right) \nonumber\\
	&=& \sum\limits_m {\left\langle f \right|{H_{e - cp - 1}}\left| m \right\rangle \left\langle m \right|{H_{e - cp - 1}}\left| i \right\rangle }/\left(E_i-E_m\right)  + \sum\limits_m {\left\langle f \right|{H_{e - cp - 2}}\left| m \right\rangle \left\langle m \right|{H_{e - cp - 2}}\left| i \right\rangle }/\left(E_i-E_m\right).
\end{eqnarray}
Note that the initial state $\left| i \right\rangle $ and final state $\left| f \right\rangle $ is
\begin{eqnarray}
	\left| i \right\rangle  &=& \left| {{n_{ks\sigma }},{n_{k's'\sigma '}},{n_{k' - q',s',\sigma '}},{n_{k + q's\sigma }};N_{ \pm q}^\nu ,N_{ \pm \left( {q - q'} \right)}^{\nu '}} \right\rangle ,\\
	\left| f \right\rangle  &=& \left| {{n_{ks\sigma }} - 1,{n_{k's'\sigma '}} - 1,{n_{k' - q',s',\sigma '}} + 1,{n_{k + q's\sigma }} + 1;N_{ \pm q}^\nu ,N_{ \pm \left( {q - q'} \right)}^{\nu '}} \right\rangle ,
\end{eqnarray}
where $n$ is the number of electron, $N$ is the number of phonon.

Lastly, the total effective Hamiltonian can be written as
\begin{eqnarray}
	{H_{eff}} &=& {H_{eff - 1}} + {H_{eff - 2}}\nonumber\\
	&=& -\frac{{{\eta ^2}}}{{{N^2}}}\sum\limits_{k,k',\sigma,\sigma',s,s'} {\sum\limits_{q,q',\nu ,\nu '} {\frac{{2\sigma \sigma '\left( {N_q^\nu  -  N_{q - q'}^{\nu '}   } \right) (\omega _q^\nu  - \omega _{q - q'}^{\nu '})}}{{{{\left( {{E _{ks\sigma }} - {E _{k + q's\sigma }}} \right)}^2} - {{(\omega _q^\nu  - \omega _{q - q'}^{\nu '})}^2}}}} } c_{k + q's\sigma }^\dag c_{k' - q's'\sigma '}^\dag {c_{k's'\sigma '}}{c_{ks\sigma }} \nonumber\\
	&&\times \left( {g_{q,q',s}^{\nu ,\nu '}g_{\left( {q - q'} \right), - q',s'}^{\nu ',\nu } + g_{ - q, - q',s'}^{\nu ,\nu '}g_{ - \left( {q - {{q'}}} \right),q',s}^{\nu ',\nu } - g_{q,q',s}^{\nu ,\nu '}g_{ - q, - q',s'}^{\nu ,\nu '} - g_{ - \left( {q - q'} \right),q's}^{\nu ',\nu }g_{\left( {q - q'} \right), - q',s'}^{\nu ',\nu }} \right)\nonumber\\
	&&- \frac{{{\eta ^2}}}{{{N^2}}}\sum\limits_{k,k',\sigma,\sigma',s,s'} {\sum\limits_{q,q',\nu ,\nu '} {\frac{{2\sigma \sigma '\left( {N_q^\nu  + N_{q - q'}^{\nu '} + 1} \right) (\omega _q^\nu  + \omega _{q - q'}^{\nu '})}}{{{{\left( {{E _{ks\sigma }} - {E _{k + q's\sigma }}} \right)}^2} - {{(\omega _q^\nu  + \omega _{q - q'}^{\nu '})}^2}}}} } c_{k + q's\sigma }^\dag c_{k' - q's'\sigma '}^\dag {c_{k's'\sigma '}}{c_{ks\sigma }}\nonumber\\
	&& \times \left( {g_{q,q',s}^{\nu ,\nu '}g_{\left( {q - q'} \right), - q',s'}^{\nu ',\nu } + g_{ - q, - q',s'}^{\nu ,\nu '}g_{ - \left( {q - q'} \right),q',s}^{\nu ',\nu } + g_{q,q',s}^{\nu ,\nu '}g_{ - q, - q',s'}^{\nu ,\nu '} + g_{ - \left( {q - q'} \right),q',s}^{\nu ',\nu }g_{\left( {q - q'} \right), - q',s'}^{\nu ',\nu }} \right).
\end{eqnarray}

According to our numerical results, the E-CP interaction provides effective repulsion for electrons, and the non-s-wave pairing symmetry could be attributed to this kind of repulsive interaction.

\section{ The derivations and expressions of $G(k)$, $\chi_0(q)$, $\chi_{s}^{zz}(q)$, $\chi_{s}^{+-}(q)$, $\chi_{c}(q)$, $\tilde{S}(q)$, $U_s$ and $\tilde{C}(q)$}

The matrix elements of the irreducible susceptibility are calculated as \cite{TakimotoPRB2004}
\begin{eqnarray}
\label{X0}
\chi_0^{a_1a_4,a_2a_3}(q)=-\frac{T}{N}\sum_k G^{a_2a_4}(k+q)G^{a_1a_3}(k).
\end{eqnarray}
$T$ is the temperature, $a_{1},a_{2},a_{3},a_{4}=1,2$. $G(k)=(ip_{n}I-M_\mathbf{k})^{-1}$ is the electron's normal Green's function  where $I$ is the unit matrix. $q=(\mathbf{q},i\omega_n)$ and $k=(\mathbf{k},ip_n)$ where $\omega_n=2n\pi T$ and $p_n=(2n-1)\pi T$ are the Matsubara frequencies with integer $n$. Within the random phase approximation (RPA), the spin and charge susceptibilities can be written as \cite{TakimotoPRB2004}
\begin{eqnarray}
\label{RPA}
\chi_{s}^{zz}(q)&=&[I-\chi_0(q)\tilde{S}(q)]^{-1}\chi_0(q),\nonumber\\
\chi_{s}^{+-}(q)&=&[I-\chi_0(q)U_s]^{-1}\chi_0(q),\nonumber\\
\chi_{c}(q)&=&[I+\chi_0(q)\tilde{C}(q)]^{-1}\chi_0(q),
\end{eqnarray}
where $\chi_{s}^{zz}(q)$ and $\chi_{s}^{+-}(q)$ represent the longitudinal and transverse spin susceptibilities, respectively, and $\chi_{c}(q)$ is the charge susceptibility. The nonzero matrix elements of $U_s$ are
$U_s^{a_1a_1,a_1a_1}=U$.
In the absence of E-CP interaction, the nonzero matrix elements of $\tilde{S}(q)$ are $\tilde{S}^{a_1a_1,a_2a_2}(q)=U\delta_{a_1a_2}$. This term can be modified as follows when the E-CP interaction is considered
\begin{widetext}
\begin{eqnarray}
\label{S}
\tilde{S}^{a_1a_1,a_2a_2}(q)&=&U\delta_{a_1a_2}-\frac{\eta^2}{N}\sum_{\mathbf{q}_1,\nu_1,\nu_1^{'}}\frac{2(N_1-N_2)(\omega_1-\omega_2)}{(\omega_1-\omega_2)^2+\omega_n^2}(g^{\nu_1,\nu_1^{'}}_{\mathbf{q}_1,-\mathbf{q},a_1} g^{\nu_1^{'},\nu_1}_{\mathbf{q}_1+\mathbf{q},\mathbf{q},a_2}+g^{\nu_1^{'},\nu_1*}_{\mathbf{q}_1+\mathbf{q},\mathbf{q},a_1}g^{\nu_1,\nu_1^{'}*}_{\mathbf{q}_1,-\mathbf{q},a_2}+2g^{\nu_1,\nu_1^{'}}_{\mathbf{q}_1,-\mathbf{q},a_1}g^{\nu_1,\nu_1^{'}*}_{\mathbf{q}_1,-\mathbf{q},a_2})\nonumber\\
&-&\frac{\eta^2}{N}\sum_{\mathbf{q}_1,\nu_1,\nu_1^{'}}\frac{2(N_1+N_2+1)(\omega_1+\omega_2)}{(\omega_1+\omega_2)^2+\omega_n^2}(g^{\nu_1,\nu_1^{'}}_{\mathbf{q}_1,-\mathbf{q},a_1} g^{\nu_1^{'},\nu_1}_{\mathbf{q}_1+\mathbf{q},\mathbf{q},a_2}+g^{\nu_1^{'},\nu_1*}_{\mathbf{q}_1+\mathbf{q},\mathbf{q},a_1}g^{\nu_1,\nu_1^{'}*}_{\mathbf{q}_1,-\mathbf{q},a_2}-2g^{\nu_1,\nu_1^{'}}_{\mathbf{q}_1,-\mathbf{q},a_1}g^{\nu_1,\nu_1^{'}*}_{\mathbf{q}_1,-\mathbf{q},a_2}),\nonumber\\
\end{eqnarray}
\end{widetext}
where $\omega_1=\omega^{\nu_1}_{\mathbf{q}_1}$, $\omega_2=\omega^{\nu_1^{'}}_{\mathbf{q}_1+\mathbf{q}}$ are the phonon dispersion relations and $N_1=\frac{1}{e^{\omega_1/T}-1}$, $N_2=\frac{1}{e^{\omega_2/T}-1}$ are the Bose distribution functions.

Finally, the nonzero matrix elements of $\tilde{C}(q)$ are
\begin{eqnarray}
\label{C}
\tilde{C}^{a_1a_1,a_2a_2}(q)&=&U\delta_{a_1a_2}-4\tilde{g}_{ep}^{2}\sum_{\nu}\frac{\omega_{\mathbf{q}}^{\nu}}{\omega_{\mathbf{q}}^{\nu2}+\omega_{n}^{2}}.
\end{eqnarray}
\section{The analysis of isotope effect}
From Eq. (S15), at $U=0$ and $\omega_n=0$, we have
\begin{widetext}
	\begin{eqnarray}
		\label{s}
		\tilde{S}^{a_1a_1,a_2a_2}(q)&=&-\frac{\eta^2}{N}\sum_{\mathbf{q}_1,\nu_1,\nu_1^{'}}\frac{2(N_1-N_2)}{\omega_1-\omega_2}(g^{\nu_1,\nu_1^{'}}_{\mathbf{q}_1,-\mathbf{q},a_1} g^{\nu_1^{'},\nu_1}_{\mathbf{q}_1+\mathbf{q},\mathbf{q},a_2}+g^{\nu_1^{'},\nu_1*}_{\mathbf{q}_1+\mathbf{q},\mathbf{q},a_1}g^{\nu_1,\nu_1^{'}*}_{\mathbf{q}_1,-\mathbf{q},a_2}+2g^{\nu_1,\nu_1^{'}}_{\mathbf{q}_1,-\mathbf{q},a_1}g^{\nu_1,\nu_1^{'}*}_{\mathbf{q}_1,-\mathbf{q},a_2})\nonumber\\
		&-&\frac{\eta^2}{N}\sum_{\mathbf{q}_1,\nu_1,\nu_1^{'}}\frac{2(N_1+N_2+1)}{\omega_1+\omega_2}(g^{\nu_1,\nu_1^{'}}_{\mathbf{q}_1,-\mathbf{q},a_1} g^{\nu_1^{'},\nu_1}_{\mathbf{q}_1+\mathbf{q},\mathbf{q},a_2}+g^{\nu_1^{'},\nu_1*}_{\mathbf{q}_1+\mathbf{q},\mathbf{q},a_1}g^{\nu_1,\nu_1^{'}*}_{\mathbf{q}_1,-\mathbf{q},a_2}-2g^{\nu_1,\nu_1^{'}}_{\mathbf{q}_1,-\mathbf{q},a_1}g^{\nu_1,\nu_1^{'}*}_{\mathbf{q}_1,-\mathbf{q},a_2}),\nonumber\\
	\end{eqnarray}
\end{widetext}
where $\omega_1=\omega^{\nu_1}_{\mathbf{q}_1}$, $\omega_2=\omega^{\nu_1^{'}}_{\mathbf{q}_1+\mathbf{q}}$ are the phonon dispersion relations and $N_1=\frac{1}{e^{\omega_1/T}-1}$, $N_2=\frac{1}{e^{\omega_2/T}-1}$ are the Bose distribution functions.

When the masses of the two sublattices are equal ($m_1=m_2=m$), if we define $\omega_1$ and $\omega_2$ as the phonon dispersion relations at $m=1$, while $\omega_{1}^{'}$ and $\omega_{2}^{'}$ are the phonon dispersion relations at $m\neq1$, then we will have
\begin{eqnarray}
	\label{omega}
	\omega_{1}^{'}&=&\omega_1/\sqrt{m},\nonumber\\
	\omega_{2}^{'}&=&\omega_2/\sqrt{m}.
\end{eqnarray}

Similarly, if we define $N_1$ and $N_2$ as the Bose distribution functions at $m=1$, while $N_{1}^{'}$ and $N_{2}^{'}$ are the Bose distribution functions at $m\neq1$, then we will have
\begin{eqnarray}
	\label{n}
	N_{1}^{'}(T)&=&\frac{1}{e^{\frac{\omega_{1}^{'}}{T}}-1}\nonumber\\
	&=&\frac{1}{e^{\frac{\omega_1}{\sqrt{m}T}}-1}\nonumber\\
	&=&N_1(\sqrt{m}T),\nonumber\\
	N_{2}^{'}(T)&=&N_2(\sqrt{m}T).
\end{eqnarray}
Finally if we define $\tilde{S}^{a_1a_1,a_2a_2}(\mathbf{q},i\omega_n=0,T)$ as the value of Eq. (\ref{s}) at $m=1$ and $\tilde{S}^{'a_1a_1,a_2a_2}(\mathbf{q},i\omega_n=0,T)$ at $m\neq1$, then we will have
\begin{widetext}
	\begin{eqnarray}
		\label{derivation}
		&&\tilde{S}^{'a_1a_1,a_2a_2}(\mathbf{q},i\omega_n=0,T)\nonumber\\&=&-\frac{\eta^2}{N}\sum_{\mathbf{q}_1,\nu_1,\nu_1^{'}}\frac{2[N_{1}^{'}(T)-N_{2}^{'}(T)]}{\omega_{1}^{'}-\omega_{2}^{'}}(g^{\nu_1,\nu_1^{'}}_{\mathbf{q}_1,-\mathbf{q},a_1} g^{\nu_1^{'},\nu_1}_{\mathbf{q}_1+\mathbf{q},\mathbf{q},a_2}+g^{\nu_1^{'},\nu_1*}_{\mathbf{q}_1+\mathbf{q},\mathbf{q},a_1}g^{\nu_1,\nu_1^{'}*}_{\mathbf{q}_1,-\mathbf{q},a_2}+2g^{\nu_1,\nu_1^{'}}_{\mathbf{q}_1,-\mathbf{q},a_1}g^{\nu_1,\nu_1^{'}*}_{\mathbf{q}_1,-\mathbf{q},a_2})\nonumber\\
		&-&\frac{\eta^2}{N}\sum_{\mathbf{q}_1,\nu_1,\nu_1^{'}}\frac{2[N_{1}^{'}(T)+N_{2}^{'}(T)+1]}{\omega_{1}^{'}+\omega_{2}^{'}}(g^{\nu_1,\nu_1^{'}}_{\mathbf{q}_1,-\mathbf{q},a_1} g^{\nu_1^{'},\nu_1}_{\mathbf{q}_1+\mathbf{q},\mathbf{q},a_2}+g^{\nu_1^{'},\nu_1*}_{\mathbf{q}_1+\mathbf{q},\mathbf{q},a_1}g^{\nu_1,\nu_1^{'}*}_{\mathbf{q}_1,-\mathbf{q},a_2}-2g^{\nu_1,\nu_1^{'}}_{\mathbf{q}_1,-\mathbf{q},a_1}g^{\nu_1,\nu_1^{'}*}_{\mathbf{q}_1,-\mathbf{q},a_2}),\nonumber\\
		&=&-\sqrt{m}\frac{\eta^2}{N}\sum_{\mathbf{q}_1,\nu_1,\nu_1^{'}}\frac{2[N_1(\sqrt{m}T)-N_2(\sqrt{m}T)]}{\omega_{1}-\omega_{2}}(g^{\nu_1,\nu_1^{'}}_{\mathbf{q}_1,-\mathbf{q},a_1} g^{\nu_1^{'},\nu_1}_{\mathbf{q}_1+\mathbf{q},\mathbf{q},a_2}+g^{\nu_1^{'},\nu_1*}_{\mathbf{q}_1+\mathbf{q},\mathbf{q},a_1}g^{\nu_1,\nu_1^{'}*}_{\mathbf{q}_1,-\mathbf{q},a_2}+2g^{\nu_1,\nu_1^{'}}_{\mathbf{q}_1,-\mathbf{q},a_1}g^{\nu_1,\nu_1^{'}*}_{\mathbf{q}_1,-\mathbf{q},a_2})\nonumber\\
		&-&\sqrt{m}\frac{\eta^2}{N}\sum_{\mathbf{q}_1,\nu_1,\nu_1^{'}}\frac{2[N_1(\sqrt{m}T)+N_2(\sqrt{m}T)+1]}{\omega_{1}+\omega_{2}}(g^{\nu_1,\nu_1^{'}}_{\mathbf{q}_1,-\mathbf{q},a_1} g^{\nu_1^{'},\nu_1}_{\mathbf{q}_1+\mathbf{q},\mathbf{q},a_2}+g^{\nu_1^{'},\nu_1*}_{\mathbf{q}_1+\mathbf{q},\mathbf{q},a_1}g^{\nu_1,\nu_1^{'}*}_{\mathbf{q}_1,-\mathbf{q},a_2}-2g^{\nu_1,\nu_1^{'}}_{\mathbf{q}_1,-\mathbf{q},a_1}g^{\nu_1,\nu_1^{'}*}_{\mathbf{q}_1,-\mathbf{q},a_2}),\nonumber\\
		&=&\sqrt{m}\tilde{S}^{a_1a_1,a_2a_2}(\mathbf{q},i\omega_n=0,\sqrt{m}T).
	\end{eqnarray}
\end{widetext}
Therefore, increasing $m$ will enhance $\tilde{S}(q)$, leading to an enlarged pairing interaction, thus an enlarged $T_c$.

In contrast, for the E-P coupling, 
\begin{eqnarray}
	\label{C}
	\tilde{C}^{a_1a_1,a_2a_2}(q)&=&-\frac{4g_{ep}^{2}}{\sqrt{m}}\sum_{\nu}\frac{\omega_{\mathbf{q}}^{\nu}}{\omega_{\mathbf{q}}^{\nu2}+\omega_{n}^{2}}.
\end{eqnarray}
At $\omega_n=0$, we define
\begin{eqnarray}
	\label{lambda_q}
	\frac{4g_{ep}^{2}}{\sqrt{m}}\frac{\omega_{\mathbf{q}}^{\nu}}{\omega_{\mathbf{q}}^{\nu2}+\omega_{n}^{2}}=\frac{4g_{ep}^{2}}{\sqrt{m}}\frac{1}{\omega_{\mathbf{q}}^{\nu}}=\lambda_{\mathbf{q}}^{\nu},
\end{eqnarray}
here $\lambda_{\mathbf{q}}^{\nu}$ is independent of $m$ since $\omega_{\mathbf{q}}^{\nu}$ scales as $\frac{1}{\sqrt{m}}$ and it cancels the $\sqrt{m}$ term in the denominator of Eq. (\ref{lambda_q}). Then we have
\begin{eqnarray}
	\tilde{C}^{a_1a_1,a_2a_2}(q)&=&-\frac{4g_{ep}^{2}}{\sqrt{m}}\sum_{\nu}\frac{\omega_{\mathbf{q}}^{\nu}}{\omega_{\mathbf{q}}^{\nu2}+\omega_{n}^{2}}\nonumber\\
	&=&-\sum_{\nu}\frac{4g_{ep}^{2}}{\sqrt{m}}\frac{1}{\omega_{\mathbf{q}}^{\nu}}\frac{1}{1+(\frac{\omega_n}{\omega_{\mathbf{q}}^{\nu}})^2}\nonumber\\
	&=&-\sum_{\nu}\frac{\lambda_{\mathbf{q}}^{\nu}}{1+(\frac{2n\pi T}{\omega_{\mathbf{q}}^{\nu}})^2}.
\end{eqnarray}
It is this relation that produces the normal isotope effect, since $\omega_{\mathbf{q}}^{\nu}(m)=\frac{1}{\sqrt{m}}\omega_{\mathbf{q}}^{\nu}(0)$, therefore $T_c(m)=\frac{1}{\sqrt{m}}T_c(0)$. It should be pointed out that the above normal isotope effect holds because $\lambda_{\mathbf{q}}^{\nu}$ is independent of $m$, otherwise it will break down, just like our E-CP case.

Compared to the E-P case, besides scaling the temperature $T$ to $\sqrt{m}T$, there is an additional $\sqrt{m}$ term as seen from Eq. (\ref{derivation}) in the E-CP case. Therefore the isotope effects are different between the E-P and E-CP cases. The latter has the inverse isotope effect.

\providecommand{\noopsort}[1]{}\providecommand{\singleletter}[1]{#1}%